\documentclass[reprint,amsmath,amssymb]{revtex4-1}
\usepackage[utf8x]{inputenc}
\usepackage{graphicx}% Include figure files
\usepackage{dcolumn}% Align table columns on decimal point
\usepackage{bm}% bold math
\usepackage{color}

\begin{document}

\preprint{APS/123-QED}

\title{Variational Monte Carlo method for the Baeriswyl wavefunction:
  application to the one-dimensional bosonic Hubbard model}

\author{B. Het\'enyi$^1$, B. Tanatar$^1$, and L.~M. Martelo$^{2,3}$}
\affiliation{$^1$ Department of
  Physics, Bilkent University, TR-06800 Bilkent, Ankara, Turkey \\
  $^2$Departamento de Engenharia F\'isica,
Faculdade de Engenharia da Universidade do Porto,
Rua Dr. Roberto Frias, 4200-465 Porto, Portugal \\
  $^3$Centro de F\'isica do Porto,
Faculdade de Ciências da Universidade do Porto,
Rua do Campo Alegre, 687, 4169-007 Porto, Portugal}

\date{\today}% It is always \today, today,

\begin{abstract}

A variational Monte Carlo method for bosonic lattice models is introduced.  
The method is based on the Baeriswyl projected wavefunction.  The Baeriswyl
wavefunction consists of a kinetic energy based projection applied to the
wavefunction at infinite interaction, and is related to the shadow
wavefunction already used in the study of continuous models of bosons.  The
wavefunction at infinite interaction, and the projector, are represented in
coordinate space, leading to an expression for expectation values which can be
evaluated via Monte Carlo sampling.  We calculate the phase diagram and other
properties of the bosonic Hubbard model.  The calculated phase diagram is in
excellent agreement with known quantum Monte Carlo results.  We also analyze
correlation functions.
\end{abstract}


\maketitle

\section{Introduction}

Variational Monte Carlo is a powerful tool to calculate the properties of
quantum systems.  In general, expectation values of physical quantities over
conveniently chosen variational wavefunctions allow the application of Monte
Carlo sampling methods.  For fermionic lattice models, commonly used
variational wavefunctions are the Gutzwiller~\cite{Gutzwiller63,Gutzwiller65}
and Baeriswyl~\cite{Baeriswyl86,Baeriswyl00,Dzierzawa97} wavefunctions (GWF
and BWF, respectively).  The GWF starts with a non-interacting wavefunction,
and projects out configurations according to the interaction.  For fermionic
systems, evaluation of physical quantities can be done approximately via a
combinatorial approximation, or exactly in the case of the
one-dimensional~\cite{Metzner87,Metzner89} and the
infinite~\cite{Metzner88,Metzner89,Metzner90} dimensional case.  In between
those two cases the state-of-the-art is the Monte Carlo method developed by
Yokoyama and Shiba~\cite{Yokoyama87a,Yokoyama87b}.  For bosonic systems, the
GWF reduces to mean-field theory~\cite{Rokhsar91}.

The BWF can be considered the counterpart of the GWF: the starting point is
the wavefunction with inifinite interaction, and the projection applied
thereonto is a function of the hopping energy.  For fermionic systems this
wavefunction already has a
history~\cite{Baeriswyl86,Baeriswyl00,Dzierzawa97,Hetenyi10,Dora16}.  For a
model of interacting spinless fermions the BWF produces excellent results for
the ground state energy~\cite{Dora16}.   We note also, that a
  method known as the momentum dependent local ansatz, in which momentum
  dependent amplitudes of pairs are used as variational parameters, was
  recently developed~\cite{Patoary13a,Patoary13b,Kakehashi14}.  While there
are a number of schemes to solve the BWF for fermionic systems, it has, to the
best of our knowledge, not been applied to bosonic systems.

In this work we develop a variational Monte Carlo (VMC) method for correlated
bosonic models based on the BWF and apply it to the bosonic Hubbard
model~\cite{Gersch63,Fisher89} (BHM) with on-site interaction.  The BHM was
originally proposed to study actual materials (bosons in porous materials),
but they were recently also realized experimentally as ultra-cold gases in
optical lattices~\cite{Jaksch98,Greiner02}.  The BHM has been treated by
analytical and numerical means, including mean-field
theory~\cite{Fisher89,Rokhsar91}, perturbative expansion~\cite{Freericks96},
quantum Monte
Carlo (QMC)~\cite{Batrouni90,Krauth91,Scalettar05,Rousseau06,Kashurnikov96b,Rombouts06}
, density matrix renormalization group
(DMRG)~\cite{Kuhner00,Ejima11,Zakrzewski08,Carrasquilla13,Kuhner98}, and exact
diagonalization (ED) ~\cite{Kashurnikov96a}.  DMRG is limited to one
dimension, QMC is limited to small system sizes, and ED is limited to even
smaller system sizes.  Our variational Monte Carlo approach is shown to give
good quantitative results, at the same time, it is not restricted to one
dimension, and is less computationally demanding than QMC or ED.  It can also
be generalized to more complex bosonic strongly correlated models with
distance dependent interaction, and/or disorder.

We calculate some of the properties of the BHM.  As expected, the ground state
energy obtained using our VMC method has a lower value than the one given by
mean-field theory.  More importantly, for the phase diagram, our results are
in excellent quantitative agreement with the quantum Monte Carlo results of
Rousseau {\it et al.}~\cite{Rousseau06,Rousseau08a,Rousseau08b}.  We also
obtain the Kosterlitz-Thouless point at the tip of the Mott lobes, and find
that our calculations underestimate the values calculated by
others~\cite{Kuhner00,Ejima11,Zakrzewski08,Kashurnikov96a,Kashurnikov96b,Rombouts06}.
We also calculate the one-particle reduced density matrix at integer and away
from integer fillings.  For integer fillings we find decay to zero.  The decay
is well approximated by an exponential function, implying the absende of a
condensate. \\

The rest of this paper is organized as follows. In the following section we
describe in detail the variational Monte Carlo method and our implementation
of it for the BHM. In section \ref{sec:results} we present our results for the
phase diagram and one-body density matrix.  Subsequently, we conclude our
work.

\section{Model and method}

\label{sec:method}

\subsection{Bosonic Hubbard model and the Baeriswyl variational wavefunction} 

We study the BHM with nearest neighbor hopping in one dimension at fixed
particle number.  The Hamiltonian is
\begin{equation}
\label{eqn:hml}
H = -J\sum_{x=1}^L \left(\hat{c}^\dagger_{x+1} \hat{c}_x + \hat{c}^\dagger_{x} \hat{c}_{x+1}\right) + U \sum_{x=1}^L \hat{n}_x (\hat{n}_x-1),
\end{equation}
where $L$ denotes the number of sites, $J$ and $U$ are the hopping and
interaction parameters, respectively.  The BWF has the form
\begin{equation}
\label{eqn:PsiBWF}
|\Psi_B\rangle = \exp(-\alpha\hat{T})|\Psi_\infty\rangle,
\end{equation}
where $\alpha$ denotes the variational parameter, and $|\Psi_\infty\rangle$ is
the wavefunction at $U=\infty$.  $\hat{T}$ denotes the hopping operator (first
term in Eq. (\ref{eqn:hml})).  The idea of the Baeriswyl wavefunction is to
start with the infinitely interacting wavefunction, and act on it with a
projector which implements hoppings.

\begin{figure}[ht]
 \centering \includegraphics[width=8cm,keepaspectratio=true]{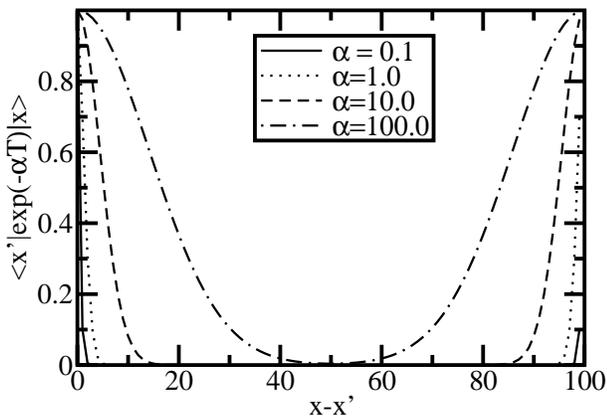}
 \caption{Kinetic energy propagator as a function of distance for different
   values of $\alpha$.}
 \label{fig:prp}
\end{figure}

\subsection{Variational Monte Carlo} 

The expectation value of an operator $\hat{O}$ can be written as
\begin{equation}
\langle \hat{O} \rangle = \langle \Psi_B|\hat{O} |\Psi_B\rangle = 
\frac{\langle \Psi_\infty |e^{-\alpha\hat{T}}
\hat{O}e^{-\alpha\hat{T}}|\Psi_\infty\rangle}{\langle \Psi_\infty |e^{-2\alpha\hat{T}}|\Psi_\infty\rangle}.
\end{equation}
The following derivations will treat a one particle system, but they are
generally applicable.  We also assume that the operator $\hat{O}$ is diagonal
in the coordinate representation.  Inserting coordinate identities, $\sum_x
|x\rangle \langle x| = 1,$ results in
\begin{equation}
\langle \hat{O} \rangle = \sum_{x_L} \sum_{x_C} \sum_{x_R}
P(x_L,x_C,x_R) O(x_C),
\end{equation}
where the probability distribution $P(x_L,x_C,x_R)$ is
\begin{equation}
P(x_L,x_C,x_R) = \frac{1}{\Omega} \langle \Psi_\infty |x_L\rangle K(x_L,x_C)
K(x_C,x_R) \langle x_R | \Psi_\infty\rangle,
\end{equation}
where
\begin{equation}
K(x,x') = \langle x |\exp(-\alpha \hat{T})|x'\rangle
\end{equation}
with $\Omega$ the normalization determined by the requirement that
\begin{equation}
\sum_{x_L,x_C,x_R} P(x_L,x_C,x_R) = 1.
\end{equation}
The quantum particle is represented by three coordinates which we call the
``left'', ``center'', and ``right'' coordinates.  Operators diagonal in the
coordinate representation can be evaluated using the center coordinate.  In
quantum Monte Carlo based methods whether finite
temperature~\cite{Pollet12,Ceperley95} or ground
state~\cite{Baroni99,Hetenyi99,Sarsa00} each particle is represented by a
large number of coordinates (Trotter slices) whose number must be increased
for accurate results as the temperature is lowered.  Therefore, while our
method is not exact as the QMC is, it is significantly less demanding of
computational resources.  Within our VMC method we can reach larger system
sizes.  In this work, we limit ourselves to system sizes with $L=50,100$, in
order to have a direct reliable comparison to the available QMC results.
\begin{figure}[ht]
 \centering
 \includegraphics[width=4cm,keepaspectratio=true]{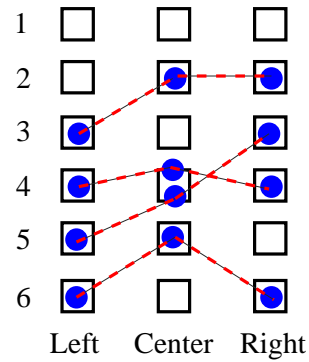}
 \caption{(Color online) Representation of our variational algorithm.  In our
   method there are three replicas of the system, labelled left, center, and
   right.  The black squares represent the lattice sites of the
   one-dimensional system for each such replica.  The blue filled circles
   represent particles.  Each particle is represented by one replica on the
   left, center, and right lattices.  The dashed red line represents the
   kinetic energy projection operator (Eq. (\ref{eqn:prpt})).  The left and
   right replicas correspond to the infinitely interacting system.  Since in
   the case there would be an infinite energy cost, there are no sites with
   more than one particle among the left and right replicas.  In the center
   replica of the lattice, two or more particles can be on the same lattice.}
 \label{fig:lcnf}
\end{figure}

The kinetic energy projection operator can be expressed as
\begin{equation}
\label{eqn:prpt}
K(x,x') = \frac{1}{L}\sum_k \exp[-\alpha \epsilon_k + i k (x-x')],
\end{equation}
and is shown in Fig. \ref{fig:prp}.  As $\alpha$ increases the propagator also
increases in value, allowing for delocalization through increased quantum
fluctuations.  Given that $P(x_L,x_C,x_R)$ is positive and normalized, MC
techniques can be applied to evaluate expectation values.  In the continuum
limit, the kinetic energy propagator reads $K(r) = I_r(2\alpha J)$ where
$I_n(x)$ are the modified Bessel functions of the first kind.

The kinetic energy can be evaluated by constructing an estimator based on
taking the logarithmic derivative of the quantity $\Omega$ with respect to the
variational parameter $\alpha$.  We can write the normalization as
\begin{equation}
\Omega = \langle \Psi_\infty |\exp(-2\alpha \hat{T})| \Psi_\infty \rangle,
\end{equation}
and the average kinetic energy as
\begin{equation}
\langle T \rangle = -\frac{1}{2 \Omega} \frac{\partial \Omega}{ \partial
  \alpha}.
\end{equation}
Writing $\Omega$ in terms of the projected wavefunction one can show that
\begin{equation}
\langle T \rangle = \sum_{x_L,x_C,x_R} P(x_L,x_C,x_R) T(x_L,x_C,x_R),
\end{equation}
where
\begin{equation}
 T(x_L,x_C,x_R) = -\frac{1}{2}
\left[
\frac{\partial \ln K(x_L,x_C)}{\partial \alpha}
+
\frac{\partial \ln K(x_C,x_R)}{\partial \alpha}
\right].
\end{equation}

The generalization to the many-body case is straightforward, but it is in
order to make some comments.  A typical configuration is represented in
Fig. \ref{fig:lcnf}.  In that figure a lattice of $6$ sites is shown.  The
left, center, and right replicas of the lattice are all represented.  A single
quantum particle is represented by three classical particles, one on each
lattice replica.  The dashed lines on the figure connecting the three
particles refer to the kinetic energy projector $K(x,x')$
(Eq. (\ref{eqn:prpt})).  Since the left and right coordinates refer to the
infinite interaction wavefunction ($\Psi_\infty$) no configuration occurs in
which more than one particle is on a particular lattice site.  This will be
the case for fillings less than one.  In general for a filling of $n$ the left
and right replicas will only have lattice sites with int($n$) or int($n$)$+1$
particles.  However, in the center replica, the lattice sites with any number
of particles can occur, since the projector does not place any restrictions
there.  {\color{red} Since} the casting of our method above is in terms of first
quantization, exchange is implemented by explicit exchange moves of pairs of
particles on the left or right lattices.  One randomly chooses a pair and then
propose the exchange as a Monte Carlo move.  This is similar to how it is done
in the continuous quantum Monte Carlo methods, such as path-integral Monte
Carlo~\cite{Ceperley95}.

\subsection{One-particle reduced density matrix}

A quantity of general interest is the one-particle reduced density matrix
(RDM).  The RDM gives information about Bose-Einstein
condensation~\cite{Penrose56}: if it tends to a finite value at long distance,
a condensate is present in the system.  The RDM (in our case for the BWF) is
given by
{\color{red}
\begin{equation}
\label{eqn:rdm}
\rho(y,x) = \langle \Psi_B |\hat{c}_y^\dagger \hat{c}_x| \Psi_B \rangle.
\end{equation}
}
The difficulty with calculating this quantity stems from the fact that it is
not diagonal in the coordinate representation.  While this is also true for
the kinetic energy, there only nearest-neighbor hoppings contribute, moreover,
one can simply take the derivative with respect to the variational parameter.
\begin{figure}[ht]
 \centering
 \includegraphics[width=4cm,keepaspectratio=true]{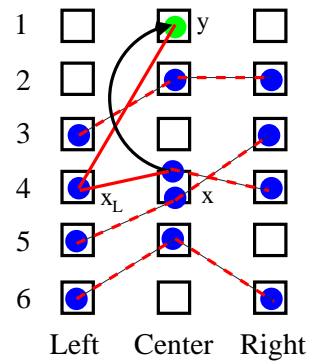}
 \caption{(Color online) Representation of a configuration with virtual
   hopping which contributes to the one-particle reduced density matrix
   (Eq. (\ref{eqn:rdm})).  The virtual hopping takes place from site $y$ to
   site $x$, indicated by the black semi-circular arrow.  The contribution to
   the average reduced density matrix is the proportion of the Baeriswyl
   propagators denoted by the straight solid red lines (see also
   Eq. (\ref{eqn:Kprop1})).  The propagators connect the left coordinate of
   the particle considered (in this case site $4$) to the central coordinate
   after virtual hopping (in this case site $1$, particle represented by green
   circle), and the propagator between the left coordinate to the unmoved
   central coordinate of the particle (site $4$, particle represented by blue
   circle). }
 \label{fig:lcnf_rdm}
\end{figure}

In the context of our variational method, the operator $\hat{c}_x^\dagger
\hat{c}_y$ corresponds to a virtual hopping from $y$ to $x$, and has the
effect of giving rise to virtual configurations in which a given particle has
two central coordinates.  One of these is located at $x$, the other at $y$.
One of these ($x$) is connected to the left coordinate of the given particle
via a Baeriswyl projector, the other ($y$) to the right coordinate.  To
calculate $\rho(x,y)$ one starts with a regular configuration, obtained from
the Monte Carlo sampling outlined above.  One choses a particle (say, with
coordinates $x_L$, $x$, $x_R$, with $x$ denoting the central coordinate) and
calculates the ratio
\begin{equation}
\label{eqn:Kprop1}
\gamma = \frac{K(x_L,y)}{K(x_L,x)}.
\end{equation}
Part of $\rho(x,y)$ is the average of contributions of this type.  The
scenario for calculating such contributions to the RDM is visually represented
in Fig. \ref{fig:lcnf_rdm}.  The two Baeriswyl projectors in the expression
in Eq. (\ref{eqn:Kprop1}) are shown in the figure by the solid straight lines.
For integer fillings, averaging over configurations of this type is all that
is needed.

\begin{figure}[ht]
 \centering
 \includegraphics[width=4cm,keepaspectratio=true]{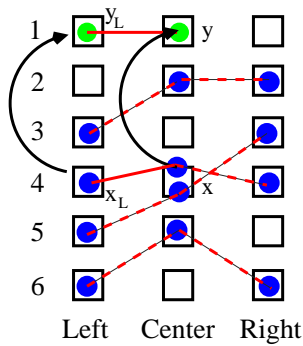}
 \caption{(Color online) Representation of a configuration with virtual
   hopping which contributes to the one-particle reduced density matrix
   (Eq. (\ref{eqn:rdm})).  The virtual hopping takes place from site $y$ to
   site $x$, indicated by the black semi-circular arrow.  At the same time, a
   virtual hopping takes place among the left coordinates, also indicated by a
   black semi-circular arrow.  The contribution to the average reduced density
   matrix is the proportion of the Baeriswyl propagators denoted by the
   straight solid red lines (see also Eq. (\ref{eqn:Kprop2})).  The
   propagators connect left coordinate of the particle (site $1$, particle
   represented by green circle) to the central coordinate of the particle
   (site $1$, particle represented by green circle), both after the virtual
   hopping, and the propagator left coordinate (site $4$, represented by a
   blue circle) and the central coordinate of the particle (site $4$,
   represented by a blue circle) both before the virtual hopping. }
 \label{fig:lcnf_rdm2}
\end{figure}

In general, there is another class of virtual configurations which needs to be
considered.  Away from integer filling, there are holes among the left and
right coordinates.  As such, the virtual hopping to which $\hat{c}_x^\dagger
\hat{c}_y$ coordsponds can also move either the left or the right coordinate.
In Fig. \ref{fig:lcnf_rdm2} this state of affairs is represented.  In this
case the quantity which must be considered is 
\begin{equation}
\label{eqn:Kprop2}
\gamma = \frac{K(y_L,y)}{K(x_L,x)},
\end{equation}
where $y_L$ represents the site to which $x_L$ is moved.  In the original
configuration, from which this virtual configuration is sampled, this site is
an empty site (or for fillings $n>1$, they are sites with int($n$) number of
particles, rather than int($n$)+1).

\subsection{Relation to shadow wavefunction}

The BWF is the lattice analog of the shadow wavefunction
(SWF)~\cite{Reatto88,Vitiello88}, used often in continuous systems in the
study of supersolidity.  To show this, we consider a one-dimensional system
with Hamiltonian
\begin{equation}
H = - \frac{\hbar^2}{2 m} \frac{\partial^2}{\partial x^2} + V(x).
\end{equation}
In one dimension the SWF is given by
\begin{equation}
\Psi(x) = \phi_1(x) \int d x'  f(|x-x'|)
\phi_2(x').
\end{equation}
The function $\phi_1(x)$ is a real-space projection operator (for now we will
take it to be one).  The term $f(|x-x'|)$ is {\it
  chosen}~\cite{Reatto88,Vitiello88} to be a Gaussian therefore we can write
\begin{equation}
\Psi(x) = A \int d x'  \exp[-C (x-x')^2] \phi_2(x'),
\end{equation}
where $A$ is the normalization and $C$ is the variational constant.  

Let us now start with a wavefunction of the form
\begin{equation}
 \exp(-\alpha \hat{T}) |\phi_2\rangle = \int d x' 
\exp(\alpha \hat{T}) |x' \rangle \langle x'|\phi_2\rangle,
\end{equation}
in which the kinetic energy propagator is applied to the state $\phi_2$.
Inserting a momentum identity, and casting the function in the coordinate
representation, results in
\begin{eqnarray}
\langle x | \exp(-\alpha \hat{T}) |\phi_2\rangle = \sqrt{\frac{m}{2 \alpha
    \pi}} \\ \nonumber \times \int d x'  
\exp\left( - \frac{m}{2 \alpha} (x-x')^2 \right)  \phi_2(x').
\end{eqnarray}
The constant $C$ is identified as $C=\frac{m}{2 \alpha}$.  The other
real-space projection $\phi_1(x)$ can be implemented also in the case of a
lattice, this would be an example of a Gutzwiller-Baeriswyl
projected wavefunction~\cite{Baeriswyl00}). 

\subsection{Implementation} 

Before MC sampling the kinetic energy propagator, as well as the estimator, is
calculated (Fig. (\ref{fig:prp})) and stored.  We apply two types of MC moves.
We move the left, central, and right coordinates by standard Metropolis
sampling from the distribution $P(x_L,x_C,x_R)$.  We also use exchange moves:
two left (or right) particles are randomly chosen and exchanged.  These moves
are {\it essential} for simulating a bosonic system.  The calculations below
show results from runs on the order of $10^6$ MC steps.  The number of
independent data points are on the order of $10^5$.  In our energy
calculations error bars typically occurred in the fourth digit of the kinetic
or potential energies.\\
\begin{figure}[ht]
 \centering
 \includegraphics[width=8cm,keepaspectratio=true]{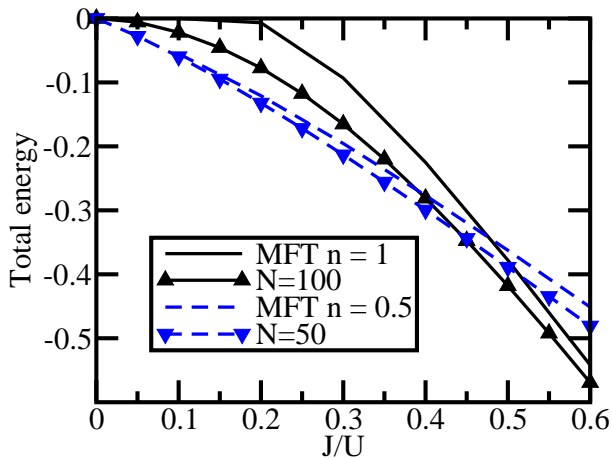}
 \caption{(Color online) Energy per lattice site for fillings of one and
   one-half calculated by our variational Monte Carlo method, and mean-field
   theory.  Lines with symbols represent variational Monte Carlo calculations,
   lines without symbols are the results of mean-field theory.  }
 \label{fig:enrg}
\end{figure}

\section{Results}

\label{sec:results}

\begin{figure}[ht]
 \centering
 \includegraphics[width=8cm,keepaspectratio=true]{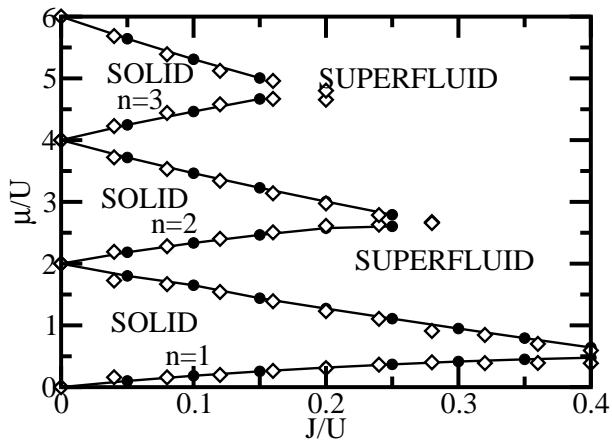}
 \caption{Phase diagram of the bosonic Hubbard model according to the
   Baeriswyl wavefunction (full circles connected by solid line) compared to
   quantum Monte Carlo Results of Rousseau {\it et al.}~\cite{Rousseau06}
   (open diamonds) recalculated via QMC~\cite{Rousseau08a,Rousseau08b}.  $n$
   denotes the filling factor.  The error bars in the calculations are smaller
   than the thickness of the symbols. }
 \label{fig:pd}
\end{figure}

For a system of $L=100$ sites we calculated the hopping and the potential
term.  The energy was minimized for different values of $J/U$.  The total
energy as a function of $J/U$ for $100$ and $50$ particles based on our
variational calculations is shown in Fig. \ref{fig:enrg}.  Also shown are
results for the same quantity from mean-field theory.  As is well-known, the
mean-field theory of the Bose-Hubbard model~\cite{Fisher89} give equations in
which the chemical potential is held fixed and the particles fluctuate.  We
solve the usual mean-field equations for a given $J/U$ adjusting the chemical
potential to correspond to an average filling of one and one-half.  The figure
shows results for the total energy without the term proportional to the
chemical potential (in order to compare the corresponding quantities from both
calculations).  The mean-field energies are quite close to the variational
Monte Carlo results, but the variational Monte Carlo results are always below
the mean-field theory.  For small $J/U$ the energy of the system with filling
one is larger than the energy for half-filling, but this changes between
$J/U=0.4$ and $J/U=0.5$.

The mean-field results indicate a phase transition at fixed filling.  At a
filling of one the phase transition occurs at $J/U \approx 0.172$, and it can
be seen in a discontinuous change in the slope of the energy and the order
parameter~\cite{Fisher89}.  In our variational calculations no discontinuity
in the slope of the energy is found, although gap closure does occur
(discussed below).  This result is qualitatively similar to what happens when
the BWF is applied to fermions: there also, no metal-insulator transition is
found~\cite{Dzierzawa97} at fixed filling. The curves of the calculated phase
diagram (Fig. \ref{fig:pd}) arise purely as a result of a phase transition
which occurs when the particle number is changed; away from integer fillings
the phase is superfluid.  \\

To calculate the phase diagram we follow the same procedure, as well as the
same parametrization, as Scalettar {\it et al.}~\cite{Scalettar05} and
Rousseau {\it et al.}~\cite{Rousseau06}.  Using the definition of the chemical
potential $\mu = E(N+1) - E(N)$ we obtain a density vs. chemical potential
curve.  The curve exhibits plateaus at integer fillings (similar to Fig. 2 of
Ref. \cite{Scalettar05}).  From the edges of the plateaus the phase diagram
can be constructed.  The results are shown in Fig. \ref{fig:pd}.  Inspite of
being a variational method, the results are in good quantitative agreement
with the quantum Monte Carlo simulations (c.f.  Fig. 11 in
Ref. \cite{Rousseau06}).  Also, for larger values of $J$ than shown in the
figure the gap closes indicating a superfluid phase.

\begin{figure}[ht]
 \centering
 \includegraphics[width=8cm,keepaspectratio=true]{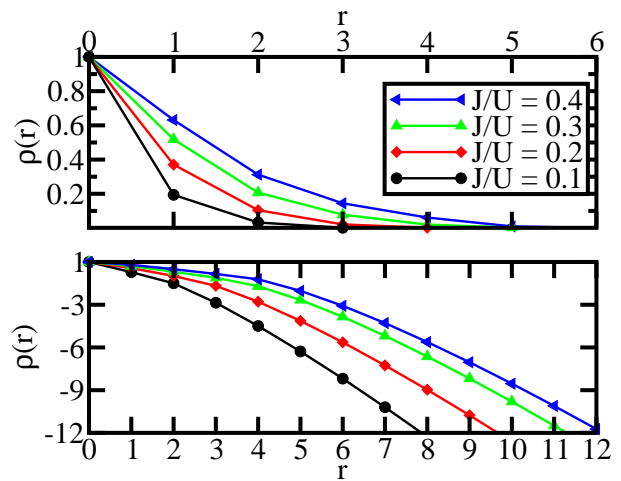}
 \caption{(Color online) One-particle reduced density matrix for systems at
   filling one with $J/U=0.1, 0.2, 0.3, 0.4$.  The bottom panel is a semi-log
   plot.  The bottom panel is a semilog plot, the $y$-axis is labelled
   according to powers of ten.  The decay of the one-body reduced density
   matrices is nearly exponential, implying the absence of a condensate.  }
 \label{fig:rdm1}
\end{figure}

At the tip of the Mott lobe, at integer filling when $J/U$ is varied a
transition is known to occur. From scaling theory it is known to belong to the
Kosterlitz-Thouless universality class~\cite{Fisher89}.  The point at which
this phase transition occurs can be estimated from inspecting the gap (it
closes at the transition point), but since small errors make a big difference
at the tip, it can also be obtained~\cite{Kuhner00} from the expression for
the gap
\begin{equation}
 \Delta(J) = A \exp\left( - \frac{B}{\sqrt{J_{KT} - J}} \right).
\end{equation}
Our results also indicate gap closure.  For a system with $L=200$ lattice
sites we obtain $J_{KT}=0.4604(2)$ from fitting this function to our data for
$J>2$, and $J_{KT}\approx 0.46$ by calculating the point where the gap closes.

The estimates given by our method significantly underestimate the
Kosterlitz-Thouless point $J_{KT}$ compared to other results in the
literature~\cite{Factor2}.  DMRG calculations of K\"uhner {\it et
  al.}~\cite{Kuhner00} find $J_{KT} = 0.594(2)$, those of Ejima {\it et
  al.}~\cite{Ejima11} find $J_{KT} = 0.610(2)$, Zakrzewski and
Delande~\cite{Zakrzewski08} find $J_{KT} = 0.5950\pm0.020$ for the first Mott
lobe, and $J_{KT} = 0.350\pm0.004$ for the second one.  An exact
diagonalization study of Kashurnikov and Svistunov~\cite{Kashurnikov96a} gives
$J_{KT} = 0.608(4)$, QMC studies find $J_{KT} =
0.600\pm0.010$~\cite{Kashurnikov96b} and $J_{KT} =
0.610(8)$~\cite{Rombouts06}.  We attribute the discrepancy between the above
results and ours to the limitation of the BWF in describing the behavior of
the system as $J$ increases.  By construction, the BWF is
expected~\cite{Baeriswyl86,Baeriswyl00} to produce reliable results for small
hoppings. 

We also calculated the RDM for several cases.  Fig. \ref{fig:rdm1} shows the
results of our calculations for a system of $L=100$ at filling one for
different values of $J/U$.  The functions show decay, although there is some
deviation from the expected exponential decay (exponential decay implies the
absence of a condensate).  Our estimates for the correlation lengths $\xi$ for
the different cases are: $\xi = 0.605(5), 0.95(3), 1.31(5), 1.66(8)$, for $J/U
= 0.1, 0.2, 0.3, 0.4$, respectively.  These results were obtained from fitting
a simple exponential function to the calculated RDMs.  We emphasize that the
exponential functions fit our correlation functions significantly better than
power law decay, as expected.

We have also calculated the RDM for systems away from integer fillings.  We
used a system of size $L=50$, with particle numbers $N=49$ and $N=48$.  In
this case, the decay does not reach zero, in other words, a finite condensate
fraction is found, which is unexpected in one dimension.  We emphasize that
our variational approach has certain limitations which are likely the cause of
this behavior.  On one hand, our $U=\infty$ function is represented in a
purely combinatorial manner, neglecting correlations between holes or extra
particles when near integer filling.  This approximation is correct in
infinite dimensions.  Apart from this, as in the original shadow wavefunction,
a spatially dependent (Gutzwiller) projector could be added to act on the
central coordinate, an approach which would improve how correlations are
captured.  This would correspond to the so-called Baeriswyl-Gutzwiller
wavefunction.

\section{Conclusion} 

\label{sec:conclusion}

We developed a variational Monte Carlo method for strongly correlated bosonic
systems based on the Baeriswyl wavefunction.  Our method was applied to the
simple bosonic Hubbard model in one dimension, but it can be generalized to
more complex models (e.g., long-range interaction, disorder), and can be
applied in any number of dimensions.  We calculated the phase diagram of the
Bose-Hubbard model, and found excellent agreement with results from quantum
Monte Carlo simulations.  Our calculations recover the shape of the Mott lobes
well.  The tip of the Mott lobes is underestimated.  We also calculated the
one-particle reduced density matrix.  At a filling of one we see decay which
is nearly exponential.  

\section*{Acknowledgments} 

We thank Nandini Trivedi and Markus Holzmann for reading our manuscript, and
making very useful comments.  We are grateful to V. G. Rousseau for providing
the results of the stochastic Green`s function calculation in
Fig. \ref{fig:pd}.  We also thank V. G. Rousseau for very useful discussions.
We acknowledge financial support from the Scientific and Technological Research Council of Turkey (TUBITAK) under grant no.s 113F334
  and 112T176.  BT acknowledges support from the Turkish Academy of Sciences (TUBA).


\begin{thebibliography}{9}
\bibitem{Factor2} The results reported here for comparison are based on a
  different definition of the Bose-Hubbard model, in which the potential is
  divided by two.  For this reason, the results reported in the original
  papers are half of the values cited here.



------------------
\end{thebibliography}
\end{document}